\documentclass[preprint,12pt]{aastex}
\usepackage{psfig}

\def\rmmat#1{{\hbox{\rm #1}}}
\def\rmscr#1{\rmmat{\scriptsize #1}}
\newcommand{\be}{\begin{equation}}
\newcommand{\ee}{\end{equation}}
\newcommand{\bt}{\begin{table} \begin{center}}
\newcommand{\et}{\end{center} \end{table}}
\newcommand{\ba}{\begin{eqnarray}}
\newcommand{\ea}{\end{eqnarray}}

\def\d{{\rm d}}

\def\eqref#1{Equation~(\ref{eq:#1})}

\begin{document}

\newcommand{\bfi}{{\bf B}} \newcommand{\efi}{{\bf E}}
\newcommand{\lel}{{\lambda_e^{\!\!\!\!-}}}
\newcommand{\me}{m_e}
\newcommand{\mcs}{{m_e c^2}}
\def\ho{{\hat {\bf o}}}
\def\hm{{\hat {\bf m}}}
\def\hx{{\hat {\bf x}}}
\def\hy{{\hat {\bf y}}}
\def\hz{{\hat {\bf z}}}
\def\hom{{\hat{\mathbf{\omega}}}}
\def\hr{{\hat {\bf r}}}
\def\omv{\mathbf{\omega}}

\title{A QED Model for the Origin of Bursts from SGRs and AXPs}

\author{Jeremy S. Heyl\altaffilmark{1,2} and Lars Hernquist\altaffilmark{1,3}}

\altaffiltext{1} {Harvard-Smithsonian Center for Astrophysics, 60
Garden St., Cambridge, MA 02138}

\altaffiltext{2} 
{Department of Physics and Astronomy; 
University of British Columbia; Vancouver, BC V6T 1Z1; Canada; 
Chandra Fellow and Canada Research Chair; heyl@physics.ubc.ca}

\altaffiltext{3} {e-mail: lars@cfa.harvard.edu }

\begin{abstract}

We propose a model to account for the bursts from soft gamma repeaters
(SGRs) and anomalous X-ray pulsars (AXPs) in which quantum
electrodynamics (QED) plays a vital role.  In our theory, that we term
``fast-mode breakdown,'' magnetohydrodynamic (MHD) waves that are
generated near the surface of a neutron star and propagate outward
through the magnetosphere will be modified by the polarization of the
vacuum.  For neutron star magnetic fields $B_\rmscr{NS} \gtrsim
B_\rmscr{QED} \approx 4.4 \times 10^{13}$~G, the interaction of the
wave fields with the vacuum produces non-linearities in fast MHD waves
that can steepen in a manner akin to the growth of hydrodynamic
shocks.  Under certain conditions, fast modes can develop field
discontinuities on scales comparable to an electron Compton
wavelength, at which point the wave energy will be dissipated through
electron-positron pair production.  We show that this process operates
if the magnetic field of the neutron star is sufficiently strong and
the ratio of the wavelength of the fast mode to its amplitude is
sufficiently small, in which case the wave energy will be efficiently
converted into an extended pair-plasma fireball.  The radiative output
from this fireball will consist of hard X-rays and soft $\gamma$-rays,
with a spectrum similar to those seen in bursts from SGRs and AXPs.
In addition, the mostly thermal radiation will be accompanied by a
high-energy tail of synchrotron emission, whose existence can be used
to test this theory.  Our model also predicts that for disturbances
with a given wavelength and amplitude, only stars with magnetic fields
above a critical threshold will experience fast-mode breakdown in
their magnetospheres.  In principle, this distinction provides an
explanation for why SGRs and AXPs exhibit burst activity while
high-field radio pulsars apparently do not.

\end{abstract}

\section{Introduction}

Soft gamma repeaters (SGRs) and anomalous X-ray pulsars (AXPs) are a
subclass of neutron stars that presently includes about a dozen
members.  They are observed to emit pulsed X-ray radiation, with
steadily increasing periods, as well as short bursts of hard X-rays
and soft $\gamma$-rays.  Compared to most radio pulsars, SGRs and AXPs
have relatively long periods that are confined to the narrow range
$P\approx 5 - 12$ s, and rapid spin-down rates, $\dot{P} \sim
10^{-11}$ s s$^{-1}$, but appear to be radio quiet.  Compared with
high-mass X-ray binaries, these objects have relatively low X-ray
luminosities, $L_x \sim 10^{35} - 10^{36}$ erg/s, and soft spectra,
but no detectable companions (for recent reviews see, e.g. Hurley
2000; Mereghetti et al. 2002).

The nature of SGRs and AXPs remains somewhat uncertain.  From timing
measurements, it is clear that they cannot be rotation-powered: their
rate of loss of rotational energy, $|\dot{E}| \equiv 4\pi^2 I
\dot{P}/P^3 \sim 10^{32.5}$, is orders of magnitude smaller than their
persistent X-ray luminosities.  Motivated by this fact, two types of
models have been proposed in which the energy is supplied primarily
either by magnetic fields or by accretion.  In the ``magnetar''
picture (Duncan \& Thompson 1992; Thompson \& Duncan 1995 [hereafter,
TD95]), SGRs and AXPs are ultramagnetized neutron stars, powered by
magnetic field decay (Thompson \& Duncan 1996; Heyl \& Kulkarni 1998),
possibly supplemented by residual thermal energy (e.g. Heyl \&
Hernquist 1997a,b).  If magnetic fields and characteristic ages can be
estimated from spin-down in the usual manner for radio pulsars
(e.g. Manchester \& Taylor 1977), then the observations imply $B_\rmscr{NS}
\sim
10^{19.5} (P \dot{P})^{1/2} \sim 10^{14} - 10^{15}$ G and $\tau_c
\equiv P/2\dot{P} \sim 10^3 - 10^5$ years (e.g. Kouveliotou et
al. 1998, 1999).  Note that these estimates would not obtain if the 
field is not a dipole (e.g. Arons 1993).\footnote{Throughout, we
use the symbol $B_\rmscr{NS}$ to denote the magnetic field of
the neutron star at its surface.}  

In the second type of theories, the X-ray emission is powered by
accretion from a low-mass or disrupted companion (e.g. Mereghetti \&
Stella 1995; van Paradijs et al. 1995; Ghosh et al. 1997), or a
fossil disk (e.g. Corbet et al. 1995; Chatterjee et al. 2000; Alpar
2001; Marsden et al. 2001).  According to these models, the magnetic
field inferred from the observed luminosities and spin periods is
$B_\rmscr{NS}\sim 10^{11}- 10^{13}$ G, if the star is accreting near its
equilibrium spin period, and the narrow observed range of spin periods
and characteristic ages can be explained by the dependence of
spin-down by the propeller effect on initial spin period and disk mass
(e.g. Chatterjee \& Hernquist 2000; Ek\c si \& Alpar 2003).

Both classes of models have strengths and weaknesses in relation to
existing observational data.  The infrared and optical counterparts
identified for some of these sources (e.g. Hulleman et al. 2000, 2001;
Wang \& Chakrabarty 2002; Israel et al. 2002, 2003a) can, in principle, be
explained by viscous dissipation and radiative reprocessing by a disk
(e.g. Perna et al. 2000; Perna \& Hernquist 2000), provided that the
disk is thin (e.g. Menou et al. 2001) and compact (see Israel et al.
2003a).  The claims of pulsed optical emission from
AXP 4U0142+61 (Kern \& Martin 2002) would appear problematic for
the accretion hypothesis.  However,
Ertan \& Cheng (2003) have shown that optical pulsations can be 
produced in the context of disk-star dynamo gap models (Cheng \&
Ruderman 1991).  It remains to be seen whether a fully consistent
solution of this type can be found which satisfies all the
requirements on e.g. the radial extent of the disk.

Tests of the magnetar model from observations in various spectral
regimes have been somewhat inconclusive.  At present, this theory does
not make detailed predictions that can be confronted with the infrared
and optical data.  Partly for this reason, it is unclear why, in this
picture, the infrared emission from AXPs appears not to be pulsed,
unlike the optical emission, at least in the case of 4U0142 (for a
discussion see, e.g. Israel et al.  2003b).  However, the recent
identification of a feature in the X-ray spectrum of SGR 1806-20
(Ibrahim et al. 2002, 2003) can be interpreted as a proton cyclotron
resonance in a magnetic field $B_\rmscr{NS} \sim 10^{15}$ G, in
agreement with timing estimates.

The recent detection of bursts from two AXPs (Gavriil et al. 2002;
Kaspi et al. 2003) has shown that these objects behave similarly to
SGRs in all respects.  Various authors have suggested that the bursts
are powered by accretion (e.g. Harwitt \& Salpeter 1973; Colgate \&
Petschek 1981; Epstein 1985; Tremaine \& Zytkow 1986; Blaes et
al. 1990; Katz et al. 1994), but Thompson \& Duncan (1996) have argued
that these models have numerous difficulties satisfying all the
observational constraints.  In particular, there is growing evidence
that the bursts are accompanied by an overall change in the properties
of the star, such as its spin-down, pulse profile, energy spectrum,
and flux (e.g. Woods et al. 2001, 2003; Kouveliotou 2003).  In some
cases, the bursts have been preceded by glitches, and it appears that
the timing noise in AXPs and SGRs is correlated with spin-down rate
(e.g. Heyl \& Hernquist 1999a; Woods et al. 2000; Gavriil \& Kaspi
2002).  These observations can be explained by a rearrangement of the
magnetic field, possibly driven by a deformation of the stellar crust
which also affects the interior of the star and its magnetosphere.
(See Ertan \& Alpar 2003 for an accretion-based interpretation of some
of these observations.)

In the magnetar model, bursts originate when a fracture of the crust
sends Alfv\'en waves into the magnetosphere (TD95).
These waves are proposed to ultimately damp, resulting in a cascade of
power to high wavenumbers, producing a pair plasma fireball through
magnetic reconnection and a burst.  This general picture is appealing
because of the observations which suggest a link between deformations of
the crust and the bursts, but the detailed physical mechanism for the
conversion of wave energy into hard X-rays and soft $\gamma$-rays is
not well-understood.  Moreover, a number of radio pulsars have been
discovered with inferred magnetic field strengths similar to those of
magnetars and apparently exceeding the value $B_\rmscr{QED} \approx
4.4 \times 10^{13}$~G (e.g. Camilo et al.  2000; Morris et al. 2002;
McLaughlin et al. 2003a,b).  It is not clear why these objects have
magnetic fields comparable to e.g. AXPs but do not exhibit AXP-like
emission.

In this paper, we propose a specific physical mechanism for the origin
of bursts in SGRs and AXPs.  Like in the TD95
theory for magnetars, we suppose that bursts are triggered by a
disturbance in the star that sends MHD waves into the magnetosphere.
However, rather than relying on magnetic reconnection to produce the
bursts, we suggest that the subsequent evolution of these waves is
driven by QED processes that, under certain conditions, causes
dissipation of the wave energy through pair production.  In our model,
vacuum polarization in fields exceeding $B_\rmscr{QED}$ leads to
non-linear evolution of fast MHD waves resembling hydrodynamic shocks.
When the fields associated with the waves develop large gradients on
scales comparable to an electron Compton wavelength, rapid conversion
of the wave energy into a pair plasma ensues, yielding
a burst.

In \S 2, we summarize the basic physics required by our analysis and
argue that QED non-linearities will affect fast-mode MHD modes, but
not Alfv\'en waves, at least to lowest order.  In \S 3, we derive an
analytical expression for the opacity of a fast mode produced at the
surface of a neutron star to develop a QED shock, and calculate the
rate at which wave energy will be dissipated into pairs.  We infer the
conditions for ``fast-mode breakdown'' to occur in neutron star
magnetospheres; i.e. when the optical depth to shock formation
approaches and exceeds unity.  We show that for a given wavelength and
amplitude of strong fast modes generated at the stellar surface, there
is a corresponding critical value of the surface magnetic field that
must be exceeded for our mechanism to operate.  Only those neutron
stars with sufficiently strong magnetic fields and which have the
capacity to produce large amplitude, short wavelength disturbances
will be susceptible to fast-mode breakdown, the development of a
pair-plasma fireball, and a subsequent burst of hard X-rays and soft
$\gamma$-rays.  In \S 4, we describe the properties of the fireball
and the nature of the emission that would be seen by a distant
observer.  We outline a scenario in \S 5 that, in principle, can
explain why SGRs and AXPs exhibit bursts, unlike radio pulsars with
intense magnetic fields.  In \S 6, we compare our model to that of
TD95, and discuss possible connections to optical and infrared
emission from AXPs.  Finally, we summarize our theory in \S 7 and
describe additional predictions that can be used to test its validity.

\section{QED Processes}

There has been considerable confusion in the recent astronomical
literature about the applicability of various results from QED to the
physics of the environments of magnetars.  In view of this, we
briefly summarize the formalism required by our analysis and discuss
regimes of validity for the relevant expressions from QED.

\subsection{The effective Lagrangian}

For our purposes, vacuum polarization effects can be calculated using
an effective Lagrangian for the electromagnetic field.  Following the
usual convention, we write
\begin{equation}
{\cal L} \, = \, {\cal L}_0 \, + \, {\cal L}_1 \, + \, \cdots \, .
\end{equation}
Here, ${\cal L}$ is the full Lagrangian density, ${\cal L}_0$ is
the classical term, and ${\cal L}_1$ includes vacuum corrections
to one-loop order.  Higher order radiative corrections would be
described by additional terms.  Dittrich \& Reuter (1985) give
an expression for the second-order correction to the Lagrangian
and find that it is smaller than the one-loop term by
a factor of the fine structure constant, regardless
of field strength.  The behavior of even higher order terms in
the expansion for ${\cal L}$ is an open question.  In what follows,
we assume that the relevant effects can be approximated using
${\cal L}_0$ and ${\cal L}_1$.

Both ${\cal L}_0$ and ${\cal L}_1$
can be expressed in terms of the Lorentz invariants
\begin{equation}
I \equiv F_{\mu \nu} F^{\mu \nu} = 2\left ( |{\bf B}|^2 \, - \,
|{\bf E}|^2 \right )
\end{equation}
and
\begin{equation}
K\equiv \left [ \epsilon^{\lambda \rho \mu \nu} 
F_{\lambda \rho} F_{\mu \nu} \right ]^2 = 
-4\left ( {\bf E} \cdot {\bf B} \right )^2 ,
\end{equation}
where $\epsilon^{\lambda \rho \mu \nu}$ is the completely 
antisymmetric Levi-Civita tensor.
The effective Lagrangian of
the electromagnetic field was derived by 
Heisenberg \& Euler (1936) and Weisskopf (1936) using electron-hole
theory.  In rationalized Gaussian units,
we can write ${\cal L}_0$ and ${\cal L}_1$ as
\begin{equation}
{\cal L}_0 \, = \, - {{1}\over 4} \, I \, ,
\end{equation}
\begin{equation}
{\cal L}_1 \, = \, {{\alpha}\over{2\pi}} \, \int_0^{\infty} \,
{\rm e}^{-\zeta} \, {{d\zeta}\over{\zeta ^3}} \, \left [
i \, \zeta^2 \, {{\sqrt{-K}}\over 4} \, {{\cos(J_+ \, \zeta) +
\cos(J_- \, \zeta)}\over {\cos(J_+ \, \zeta) - \cos(J_- \, \zeta)}}
\, + \, B_\rmscr{QED}^2 \, + \, I \, {{\zeta^2}\over 6} \, \right ] ,
\end{equation}
where
\begin{equation}
J_{\pm} \, \equiv \, {1\over{2 B_\rmscr{QED}}} \, \left [
-I \, \pm \, i \, \sqrt{-K} \right ]^{1/2} ,
\end{equation}
$\alpha \equiv e^2/\hbar c$ is the fine structure constant,
$B_\rmscr{QED} \equiv m^2c^3/e\hbar \approx 4.4\times 10^{13}$ G, and
a similar quantity can be defined for the electric field,
$E_\rmscr{QED} \equiv m^2c^3/e\hbar \approx 2.2\times 10^{15}$ V/cm.
The above expressions for ${\cal L}_0$ and ${\cal L}_1$ are identical
to the corresponding terms in eq. (45a) of Heisenberg \& Euler (1936).

The above integral cannot be evaluated explicitly, in general.  Heyl
\& Hernquist (1997c; hereafter HH97) have derived an analytic
expression for ${\cal L}_1$ as a power series in $K$:
\begin{equation}
{\cal L}_1 \, = \, {\cal L}_1 (I,0) \, + \, K \, {{\partial {\cal L}_1}
\over{\partial K}} \Bigg |_{K=0} \, + \, \cdots \, ,
\label{eqnL}
\end{equation}
where the first term in this series is
\begin{equation}
{\cal L}_1 (I,0) \, = \, {\alpha\over{4\pi}} \, I \, X_0 \left ( 
{1\over \xi} \right ) \, = \, {\alpha\over{4\pi}} \,
\int_0^\infty \, {\rm e}^{-u/\xi} \, {{du}\over{u^3}} \,
\left ( -u \coth u \, + \, 1 \, + \, {{u^2}\over 3} \right ) ,
\end{equation}
and
\begin{equation}
\xi \, = \, {1\over B_\rmscr{QED}} \, \sqrt{{I\over2}} \, .
\end{equation}
The function $X_0$ can be evaluated analytically (as can the
higher order terms in the expansion for ${\cal L}_1$) with
the result (HH97)
\begin{equation}
X_0(x) \, = \, 4\int_0^{x/2-1} \, \ln (\Gamma(v+1)) \, dv \, - \,
{1\over 3} \, \ln x \, + \, {\cal C} \, - \, \left [ 
1 \, + \, \ln \left ( {{4\pi}\over x} \right ) 
\right ] \, x \, + \, \left [ {3\over 4} \,
+ \, {1\over 2} \, \ln \left ( {2 \over x} \right ) \right ] \, x^2 \, ,
\end{equation}
where
\begin{equation}
{\cal C} \, = \, 2\ln 4\pi \, - \, 4\ln A \, - {5\over 3} \ln 2 \, = \, 
2.911785285,
\end{equation}
and
the constant $\ln A$ is related to the first derivative of the
Riemann zeta function, $\zeta^{(1)}(x)$, by
\begin{equation}
\ln A \, = \, {1\over {12}} \, - \, \zeta^{(1)} (-1) \, = \,
0.248754477 \, .
\end{equation}
The integral of $\ln \Gamma (x)$ can be expressed in terms of
special functions (eqs. 18, 19 in HH97).  (See also Dittrich 
et al. 1979; Ivanov 1992, but note the cautionary remark
in HH97.)

The expression above for $X_0(x)$ can be expanded in either
a Taylor series in the weak field limit, $\xi \ll 1$, or an
asymptotic series in the strong field limit $\xi \gg 1$, as
can the higher order terms in equation (\ref{eqnL}), to give
series expansions for ${\cal L}_1$ as a function of either $I$ and
$K$, or equivalently $B$ and $E$.  In particular, to lowest
order in the weak field limit
\begin{equation}
{\cal L}_1 \, = \, {{\alpha}\over{90 \pi}} \, {1\over {B_\rmscr{QED}^2}}
\left [ (B^2 \, - \, E^2)^2 \, + \, 7 ({\bf E} \cdot {\bf B})^2 
\right ] \, + \, \cdots \,\,\, (\xi \ll 1) \, .
\label{HEL1}
\end{equation}
In the limit of an ultrastrong magnetic field, $B \gg B_\rmscr{QED}$,
but for a weak electric field, $E \ll E_\rmscr{QED}$, 
${\cal L}_1$ can be written
\begin{equation}
{\cal L}_1 \, = \, {{\alpha}\over{6 \pi}} \, B^2 \,
\left [ \ln \left ( {B\over {B_\rmscr{QED}}} \right ) \, - \, 12 \ln A
\, + \, \ln 2 \right ] \, + \, \cdots \, 
\label{HEAL1}
\end{equation}
(see e.g. eq. 29 in HH97 for the higher order terms).  We note that
equations (\ref{HEL1}) and (\ref{HEAL1}) agree, respectively,
with the corresponding terms in eqs. (43) and (44) of
Heisenberg \& Euler (1936).  The analysis of HH97 generalizes the
expressions of Heisenberg \& Euler to arbitrary order.

Also, note that while our expression for ${\cal L}_0$ is identical 
to eq. (4-120) of Itzykson \& Zuber (1980), equation (\ref{HEL1}) 
differs from their eq. (4-125) by a factor of 
$1/4\pi$, as a consequence of
a difference in the system of units employed.\footnote{Itzykson
\& Zuber (1980) use Heaviside's units in defining the Coulomb
force; $E^2$ and $B^2$ are smaller in this system than in ours by
a factor of $4\pi$.}  Our expressions for 
${\cal L}_0$ and ${\cal L}_1$ both differ from those in
Berestetskii et al. (1982) by an overall factor of $1/4\pi$ 
(their eqs. 129.2 and 129.21); however, the dynamics of the fields
are invariant with respect to rescalings of the Lagrangian.

We emphasize that the expression for the Lagrangian in the weak-field
limit, equation (\ref{HEL1}), cannot be applied to magnetar fields
which are thought to have $B_\rmscr{NS} \gg B_\rmscr{QED}$.  The use of the
weak-field expressions to calculate, e.g. the index of refraction of
the vacuum near the surface of a magnetar will result in estimates
that are incorrect by more than an order of magnitude at the relevant
field strengths.  In this limit, the Lagrangian should instead be
approximated by e.g. equation (\ref{HEAL1}), which is an asymptotic
series for ${\cal L}_1$ valid for $B \gg B_\rmscr{QED}$ and $E \ll
E_\rmscr{QED}$.

\subsection{The magnetized vacuum}

The dependence of ${\cal L}_1$ on $B$ and $E$ means that the vacuum
will behave in some respects like a material medium in the presence of
strong magnetic and/or electric fields.  This behavior will, in
general, be more complex than for simple polarizable media
because of the non-linear dependence of ${\cal L}_1$ on the
fields (e.g. Erber 1966; Tsai \& Erber 1975).

Many of these effects are subtle even for magnetar field strengths.
For example, the vacuum will respond to an applied field like a
non-linear paramagnetic substance (e.g. Mielnieczuk et al.  1988), so
that $B$ and $H$ will differ (Klein \& Nigam 1964a,b; Heyl \&
Hernquist 1997d).  This will influence e.g.\ the spin-down rate of
magnetars, but quantitatively the effect appears to be small
(Heyl \& Hernquist 1997e).

More promising is the impact that vacuum polarization can have on the
propagation of waves through the magnetospheres of strongly magnetized
neutron stars.  Because a magnetized vacuum responds to applied fields
in a non-linear manner, sinusoidal disturbances like electromagnetic
waves can develop discontinuities that are analogous to hydrodynamical
shocks.  Lutzky \& Toll (1959) and Zheleznyakov \& Fabrikant (1982)
studied this process in the weak-field limit, while Bialynicka-Birula
(1981) considered stronger fields and Heyl \& Hernquist (1998a;
hereafter HH98) extended the analysis to fields characteristic of
magnetars.  In particular, HH98 used the effective Lagrangian
described in \S 2.1 to derive an expression for the opacity of
electromagnetic waves traveling through an arbitrarily strong magnetic
field to experience shocks, used the method of characteristics to
investigate the development on discontinuities in these waves and
investigated the jump conditions across these discontinuities (see
also Boillat 1972).

Similar considerations apply to certain types of MHD waves.  Heyl \&
Hernquist (1999b; hereafter HH99) employed a variational principle to
investigate the propagation of MHD waves through strongly magnetized,
relativistic plasmas, including QED effects (see Achterberg 1983 for
the classical limit).  HH99 showed that the fundamental modes of
propagation in the ultrarelativistic limit are two oppositely directed
Alfv\'en modes and the fast mode, in agreement with Thompson \& Blaes 
(1998, hereafter TB98).

In the ultrarelativistic limit, the fast mode does not carry any
current and behaves similar to a vacuum electromagnetic wave (TB98);
consequently, it is not surprising that the fast MHD modes are subject
to shocking in a manner identical to electromagnetic radiation
traveling through an external magnetic field in the absence of a
plasma (HH99).  The ultrarelativistic limit results from neglecting
the mass of the charge carriers; therefore, it corresponds to the
long-wavelength limit and applies to waves whose frequency is well below 
the plasma frequency (TB98).

It is, thus, these long-wavelength fast modes that are susceptible to
energy dissipation through pair production, at least to lowest order.
On the other hand, HH99 found that because of the gauge and Lorentz
invariance of QED, non-linearities do not develop in the propagation
of single Alfv\'en modes.

\section{Fast-Mode Breakdown}

In a manner analogous to the general picture proposed by TD95, we
suppose that bursts are triggered in SGRs and AXPs either by a
fracture of the stellar crust or a pure rearrangement of the magnetic
field.  Shifts in the footprints of magnetic field lines will send MHD
waves into the magnetosphere (e.g. fig. 1 in TD95).  For the purposes
of discussion, we assume that some fraction of the wave energy will be
emitted in the fast mode.  Whether the fast modes are produced
directly or through the interaction of Alfv\'en waves is beyond
the scope of this article.    Unlike the scenario described in TD95, we
do not consider the evolution of Alfv\'en waves since the analysis of
HH99 indicates that they will not develop shocks from their
interaction with the vacuum.  (QED will introduce additional couplings
between oppositely directed Alfv\'en waves, perhaps exciting other
dissipative processes, but we do not pursue these interesting
possibilities here.)

As we noted earlier, the evolution of the fast mode in a strongly
magnetized, ultrarelativistic plasma is identical to the behavior of
an electromagnetic wave in the absence of the plasma; therefore, we
use the results of HH98 to describe the propagation of a fast wave
through the magnetosphere of a neutron star.  

To illustrate the effect, we consider
waves in which the electric field is polarized perpendicular to the
magnetic field of the star, and require the wave to travel
perpendicular to the magnetic field.  Under these assumptions, the
opacity for a fast mode of wavenumber $k$ to form a shock is given by
\begin{equation}
\kappa \, = \, - k \, \frac{\alpha}{4\pi} \, \frac{b}{B_\rmscr{QED}} \,
\left [ 
X_0^{(1)} \left ( \frac{1}{\xi} \right ) \xi^{-2} -
X_0^{(2)} \left ( \frac{1}{\xi} \right ) \xi^{-3} +
X_0^{(3)} \left ( \frac{1}{\xi} \right ) \xi^{-4} \right ] ,
\end{equation}
where $\xi=B_\rmscr{NS}/
B_\rmscr{QED}$, $B_\rmscr{NS}$ is the strength of the stellar field, 
$b$ is the amplitude of the magnetic field of the fast mode, 
the function $X_0(x)$ is defined by equations (10-12),
and $X_0^{(n)}(x) \equiv \frac{\d^n}{\d x^n} X_0(x)$.
Using the definition of $X_0(x)$, we find
\begin{equation}
\kappa \, = \, k \,\frac{b}{B_\rmscr{QED}} \, \frac{\alpha}{\pi} \,\left \{ 
\frac{1}{3 \xi} \, + \,
\frac{1}{\xi^2} \left [ \frac{1}{4} \ln \left ( 4\pi\xi \right ) - \frac{1}{2} \ln \Gamma \left (\frac{1}{2\xi} \right ) + \frac{1}{2} \right ] 
+ \frac{1}{4\xi^3} \Psi \left (\frac{1}{2\xi} \right ) 
- \frac{1}{8\xi^4} \Psi^{(1)} \left (\frac{1}{2\xi} \right ) 
\right \},
\label{eqkappa}
\end{equation}
where $\Psi (x) \equiv d\ln \Gamma / dx$, and $\Psi^{(1)}(x) =
d\Psi/dx$.

HH98 give expansions for $\kappa$ that are valid in the weak- and
strong-field limits and which are useful for determining the
asymptotic behavior of the opacity.  In what follows, we are
interested in the evolution of a wave emitted from the stellar
surface, where the field would be stronger than $B_\rmscr{QED}$
for a magnetar, as it propagates into regions in the magnetosphere
where $B \ll B_\rmscr{QED}$, requiring us to use the full
expression for $\kappa$ above.

Note that according to equation (\ref{eqkappa}), the opacity, and
hence the integrated optical depth, depend on the properties of the
wave only through the combination $k b / B_\rmscr{QED}$; i.e. the
wavenumber ($2\pi/\lambda$) multiplied by the amplitude in units of
$B_\rmscr{QED}$.

The structure of the wave at a particular location is determined by
integrating the opacity over the trajectory of the wave, giving the
optical depth, $\tau = \int \kappa \d s$.  We assume that the wave is
initially sinusoidal; therefore, the wave begins to lose energy to pair
production when $\tau$ reaches unity.  We use the Maxwell-equal-area
prescription (Landau \& Lifshitz 1987) to determine the structure of the wave
after the shock forms and the ensuing rate of pair production.

In this way, we adopt a number of simplifying assumptions to describe
the rate of energy dissipation.  Strictly speaking, the effective
Lagrangian we employ here is correct only if the magnetic field is
locally uniform.  This description is no longer valid when the field
varies by of order unity on scales comparable to an electron Compton
wavelength.  Analyzing this situation would require a more general
formulation, such as the proper-time method developed by Schwinger
(1951).  In practice, this means that we cannot describe the structure
of the shocks.  However, the shock jump conditions (HH98) make it
possible to estimate the rate of dissipation through pair production
without a detailed understanding of shock structure.

We also assume that the only dissipative mechanism is pair production.
In particular, we ignore dispersive effects, as they are estimated to
be small under the conditions appropriate for neutron star
magnetospheres (e.g. Adler 1971).  In addition, for simplicity, we do
not include other processes, such as non-linear QED interactions 
between Alfv\'en waves.

\begin{figure}
\plotone{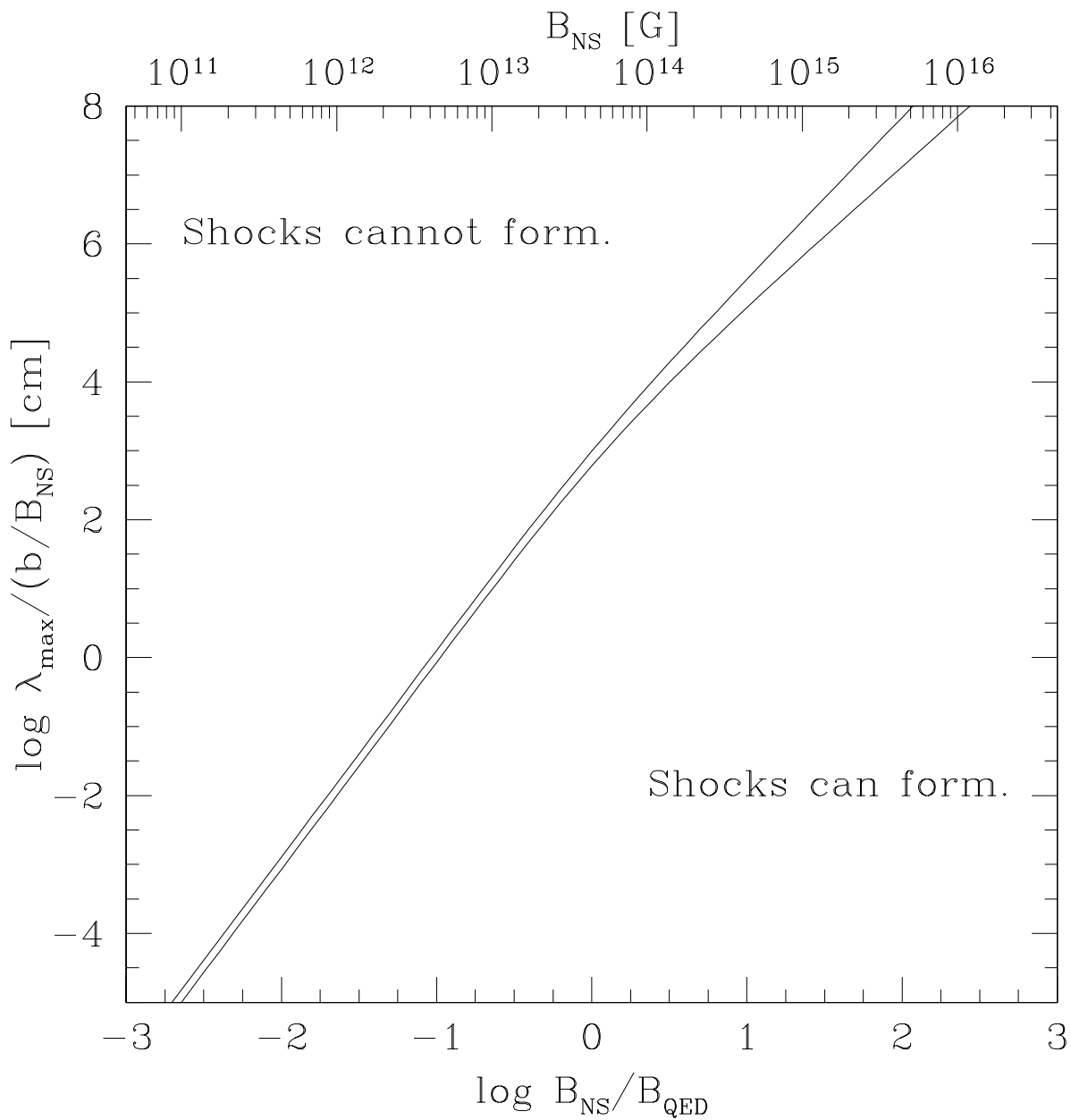}
\caption{The conditions for a fast mode to break down in a
neutron star magnetosphere, expressed as a ratio of the
wavelength to the amplitude of the wave.  Only waves with
sufficiently small wavelength and/or large amplitude
will break down.
The calculation
for the upper curve assumes that the wave is plane parallel (i.e. its 
strength does not diminish with radius from the star).  The lower curve 
is for a spherical wave.}
\label{fig:wavelength}
\end{figure}

To summarize, we assume that a sinusoidal fast mode is generated at
the stellar surface and that it begins to steepen owing to vacuum
polarization as it propagates outward (see fig. 4 in HH98).  If the
integrated optical depth exceeds unity along its trajectory, the field
gradients in the wave will be large enough for pairs to be produced,
leading to ``fast-mode breakdown.''

Figure~\ref{fig:wavelength} depicts the conditions for a fast mode to
break down in this manner in the magnetosphere of a neutron star with
a given surface field.  Here, we assume, that the fast mode is
polarized with its electric field perpendicular to the stellar field,
that the wave travels radially in the equatorial plane of the star,
whose radius is ten kilometers, and that the neutron star field is a
pure dipole.  For simplicity, we have not included the gravitational
redshift of the wave which would decrease the critical wavelength by
at most 30\%.  Although the shape of the fast modes with wavelengths
larger and amplitudes smaller than traced by the lines will change as
they propagate through the magnetosphere, they do not form shocks, and
so no energy is dissipated.

If the ratio of the wavelength to amplitude of strong fast modes
generated at the surface of the neutron star is limited from below,
then Figure~\ref{fig:wavelength} demonstrates that only stars with
surface fields greater than a particular value will support fast-mode
breakdown in their magnetospheres.  For example, if the minimum
wavelength of a fast mode is one kilometer, one finds that only
neutron stars with $B_\rmscr{NS} > 10 B_\rmscr{QED}$ will exhibit
fast-mode breakdown for disturbances with amplitude $b= B_\rmscr{NS}$.
Our model thus naturally leads to a cut-off behavior in the production
of a pair plasma through this mechanism, depending on the manner in
which waves are produced at the stellar surface.

\begin{figure}
\plottwo{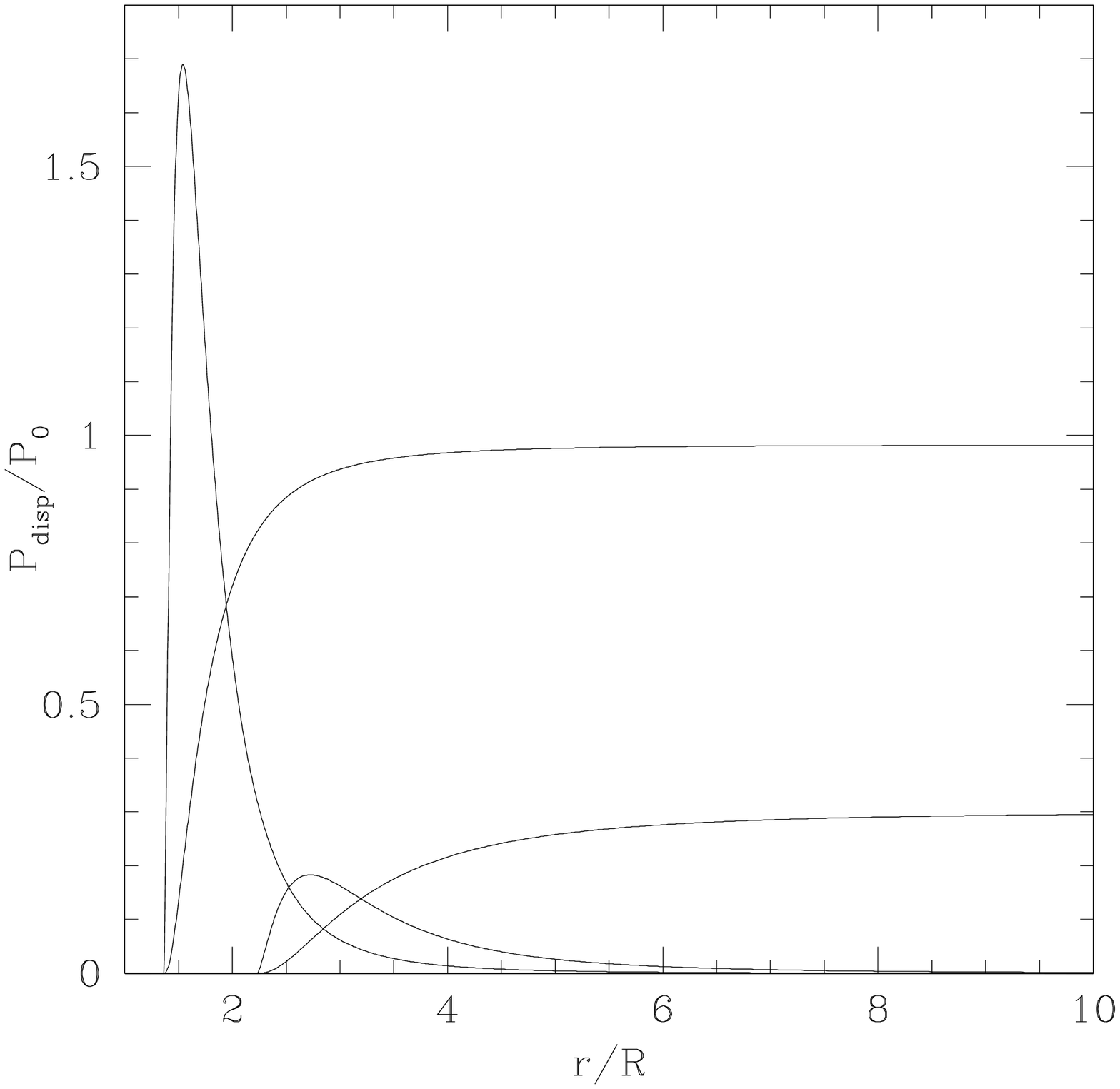}{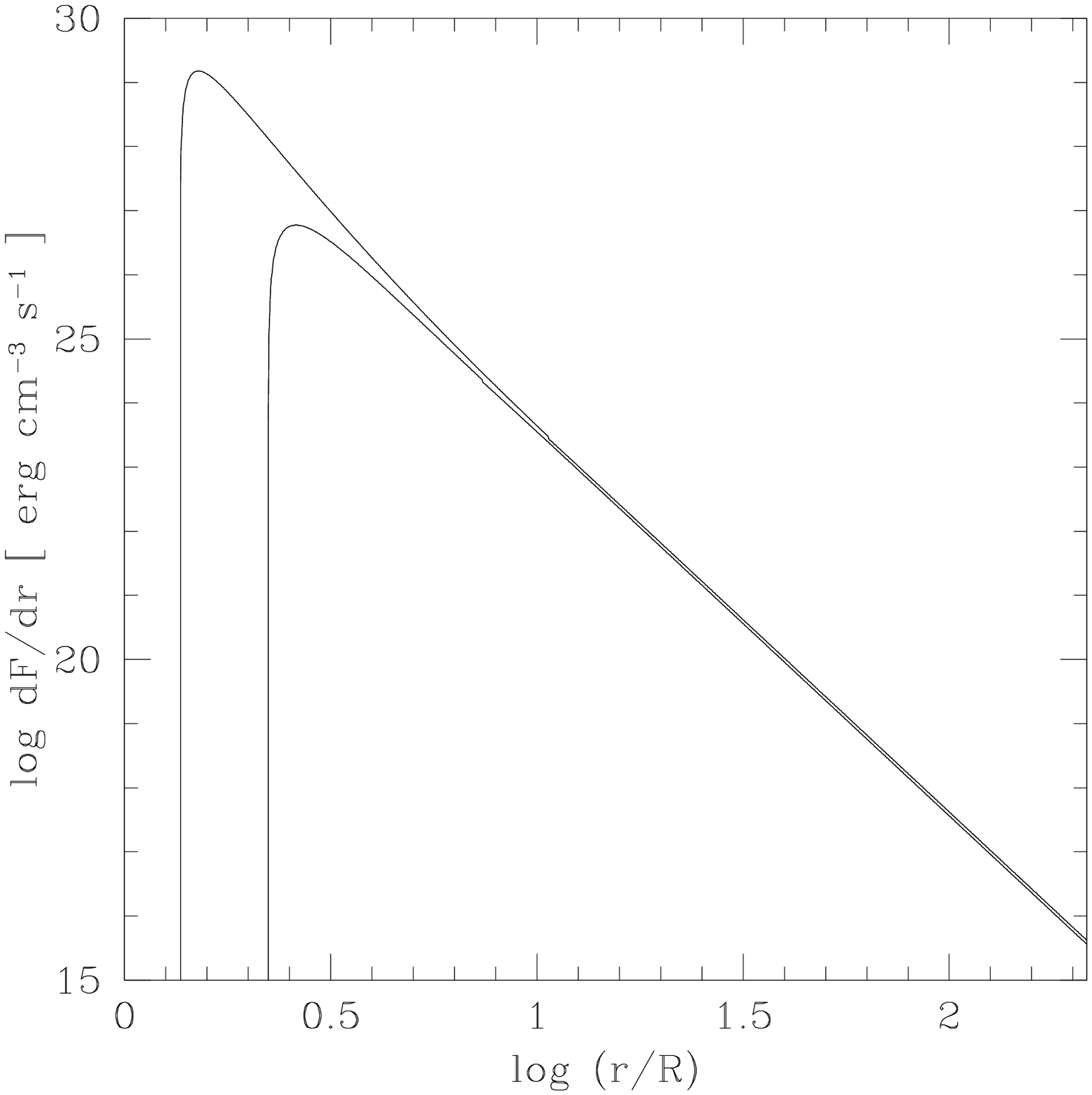}
\caption{Rate of power dissipation of a fast-mode traveling through a 
magnetar magnetosphere.  The initial amplitude of the fast mode is 1\% of 
the surface magnetic field, and its wavelength is 6.3 meters.  Similar
results obtain as the amplitude and wavelength are increased 
proportionately.  The absolute power dissipated increases as the square of
the wave amplitude.   In the left panel, we show the cumulative (monotonic
curves) and differential (peaked curves) energy dissipated for stellar 
fields $B_{NS}=30 B_\rmscr{QED} \approx 1.2 \times 10^{15}$~G (upper curves)
and $B_{NS}=10 B_\rmscr{QED} \approx 4.4 \times 10^{14}$~G (lower curves).
The right panel shows the energy production rate for the two cases, with
the upper and lower curves for 
$B_{NS}=30 B_\rmscr{QED}$ and $B_{NS}=10 B_\rmscr{QED}$, respectively.
}
\label{fig:power}
\end{figure}

In Figure~\ref{fig:power}, we focus on the evolution of a particular
wave in greater detail.  Here, we examine a fast mode of six
meter wavelength (to be precise, $2\pi R/10^4$) whose initial amplitude
is equal to one percent of the surface magnetic field.  We consider
surface fields of $4.4 \times 10^{14}$~G and $1.2 \times 10^{15}$~G.
The left panel shows both the cumulative fraction of wave energy that
is converted into pairs, along with the differential fraction per
radial bin.  For the more strongly magnetized neutron star, the bulk
of the energy of the wave dissipates within one stellar radius from
the surface.  Only about 30\% of the energy in the wave traveling
through the magnetosphere of the more weakly magnetized neutron star
is dissipated.  Again, the bulk of the pair production is within a few
stellar radii.  The right panel depicts the energy production rate for
the two situations.

\section{The Fireball}

In our model, most of the pairs are produced near the surface of the
star, within a few stellar radii.  To place the energy generation
rates in perspective, the active volume of pair production is $\sim
10^{19}$cm$^3$, yielding a total energy generation rate of $\sim
10^{47}$erg/s for the more strongly magnetized star in Figure 2.  The
typical energy of an SGR burst is $\sim 10^{40}$~ergs, so the
pair-production will last $\sim 100$ ns or a few oscillations of the
fast mode.

\begin{figure}
\plotone{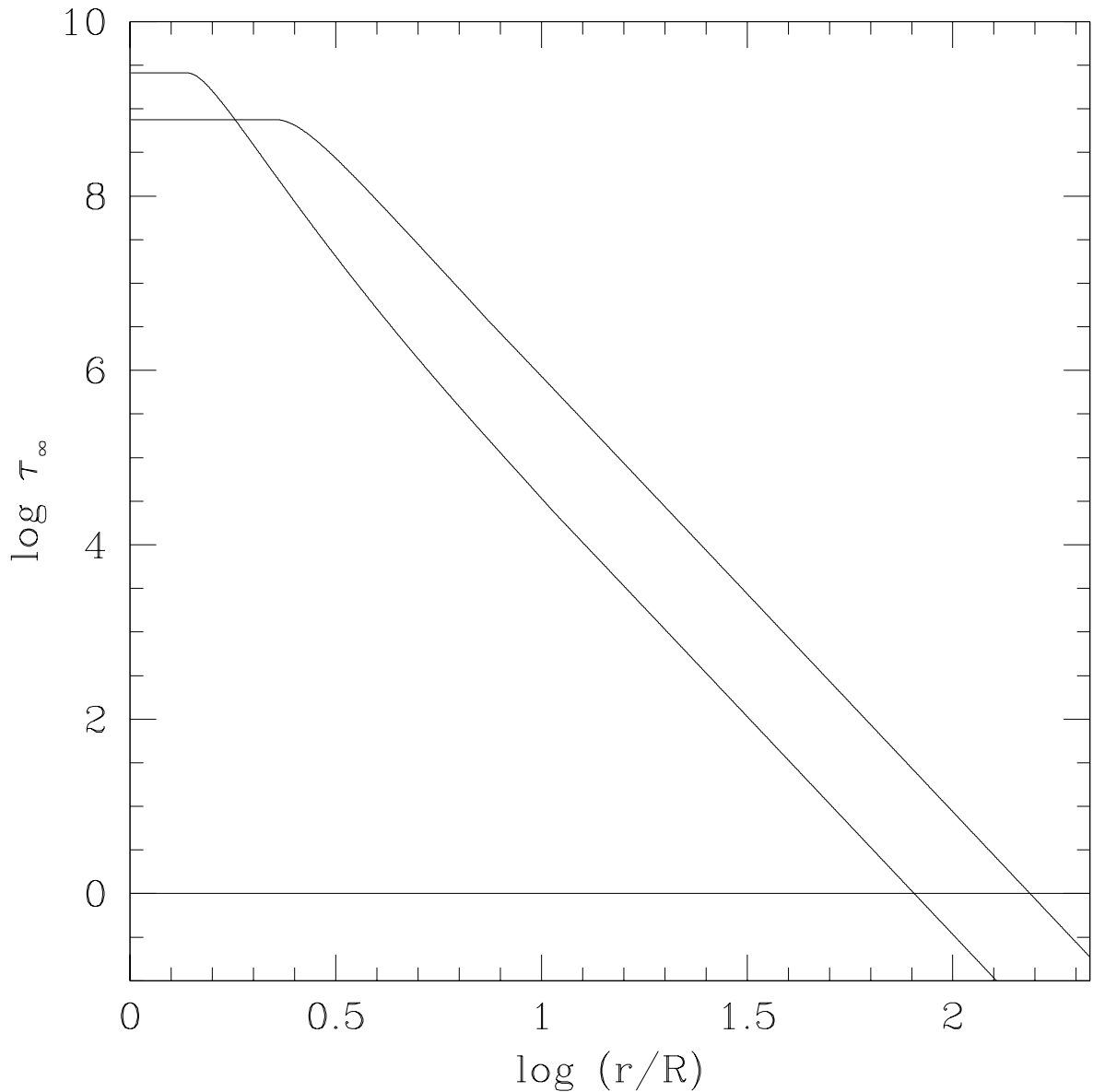}
\caption{The Thomson optical depth for X-rays to escape from a given
radius to infinity for the models depicted in Fig. 2.  To calculate
the density of pairs, we have assumed that the fireball subtends one
steradian, the total energy of the burst is $10^{40}$~ergs and that
this energy is initially in the form of the rest-mass energy of the
pairs.  The stronger field case has a larger peak optical depth $\sim
10^9$ because the pair production in the strong field is typically
closer to the star.  }
\label{fig:tau}
\end{figure}

However, the total volume of the fireball is much larger than this
because each electron has a large cross section to absorb X-rays.  A
mere $10^{13}$ ergs of pairs per cubic centimeter is sufficient to
make a distance of several kilometers opaque to X-rays.
Figure~\ref{fig:tau} depicts the Thomson optical depth for X-rays to
escape to infinity radially from various distances from the center of
the neutron star.  The fireball extends for several tens or hundreds
of stellar radii from the surface of the neutron star for the models
depicted in Figures~\ref{fig:power} and~\ref{fig:tau}.  Additionally,
as long as the fast mode is not discontinuous initially, no pairs are
produced at the surface of the star.  This gap between the fireball
and the stellar surface prevents a large baryon contamination of the
pair plasma.

The size of the fireball depends on the properties of the wave,
specifically on $k b$, the product of the wavenumber and the amplitude
of the wave.  Using the asymptotic behavior of the shock opacity for
weak fields (HH98) and the evolution of the wave for large values of
$\tau$, we find that the outer radius of the fireball goes as $(k
b)^{-2/5}$.  The distance between the surface of the star and the edge
of the fireball is proportional to $(k b)^{-1}$.  To make things a bit
more concrete, for a neutron star with surface dipole field of
$B_{NS}=30 B_\rmscr{QED}$, the initial fireball extends from 4~km
above the surface of the 10~km star to 800~km from the center of the
star for $k b= 0.3 B_\rmscr{QED}$m$^{-1}$.  If we look at a wave with
$k=1$m$^{-1}$ and $b=B_{NS}$ ($k b=30 B_\rmscr{QED}$m$^{-1}$), the
fireball starts a mere 40~m above the surface and extends to 120~km
from the star.  Because the pair production begins only after the wave
has traveled one optical depth, there is always a gap between the
stellar surface and the fireball.

The formation of the fireball and the fireball itself will produce two
types of emission.  The vast majority of this emission will be in the
form of blackbody radiation from the expanding fireball (as in TD95).
The model presented here differs in the mechanism that produces the
pair-plasma fireball but the fireball itself is qualitatively similar,
so the specific radiative processes outlined by TD95 operate here as
well.  We can obtain an estimate for the effective temperature of the
blackbody emission by assuming that the fireball emits at the
Eddington limit in the strong magnetic field surrounding the neutron
star.  TD95 give the following result (their eq. 44):
\begin{equation}
T_\rmscr{max} (\|) = 7.6 (Y_{e^-} + Y_{e^+} )^{-1/6} \left [ \frac{B(R)}{B_\rmscr{QED}} \right ]^{1/3} \left ( \frac{R}{R_{NS}} \right )^{-1/3} \left ( 
\frac{g_{NS,14}}{2} \right )^{1/6} \rmmat{keV}.
\end{equation}
Fast-mode breakdown produces a fireball many tens of stellar radii in
diameter, so the value of $R$ at the outside of the fireball is much
greater than the radius of the star.  However, the bulk of the energy
of the fireball is dumped within a few radii of the star and the
fireball is contained within the closed field lines; therefore,
emission can escape from near the surface of the star, and the bulk of
the flux will be characterized by a blackbody temperature $\sim
10$~keV.

\begin{figure}
\plottwo{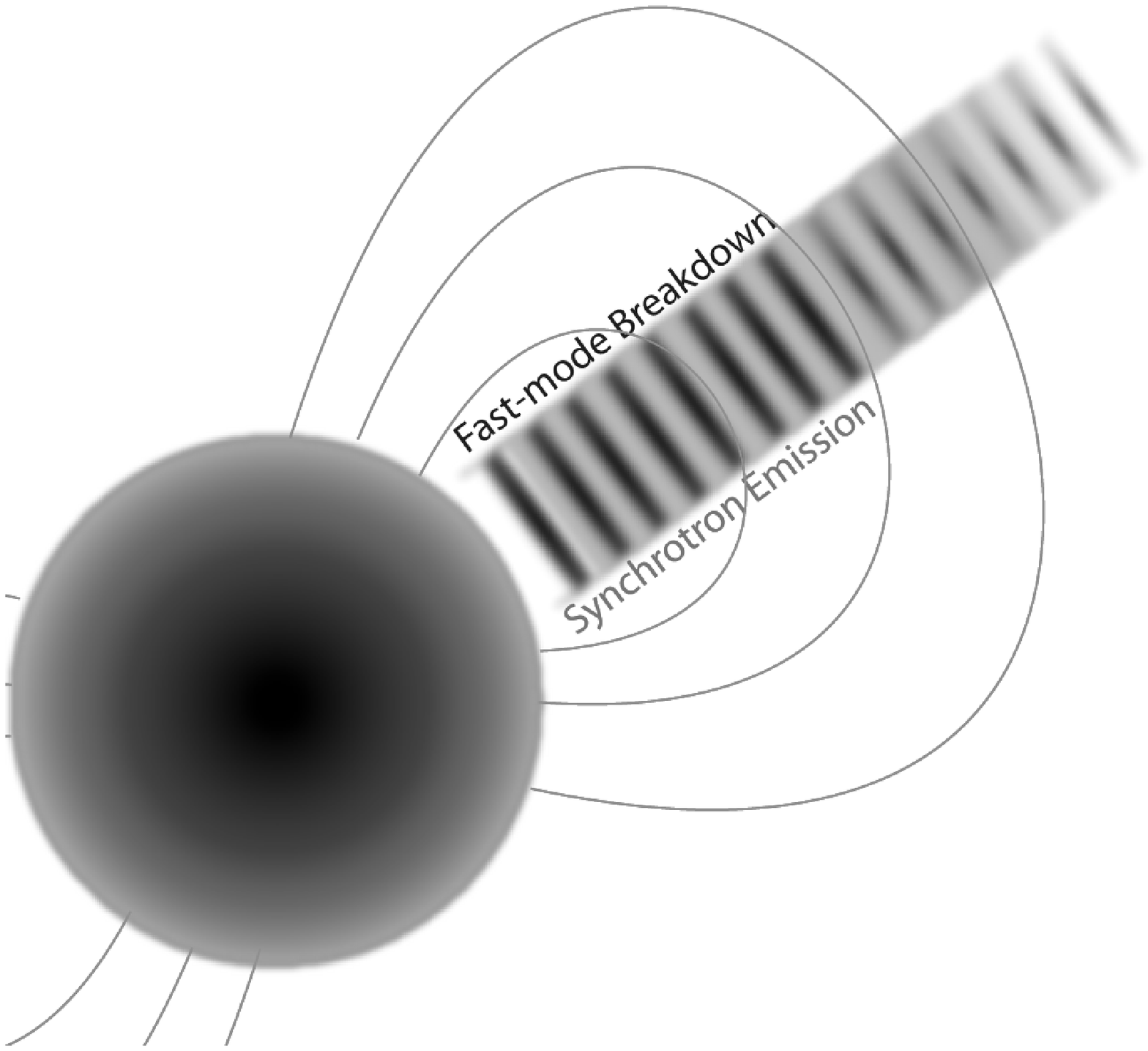}{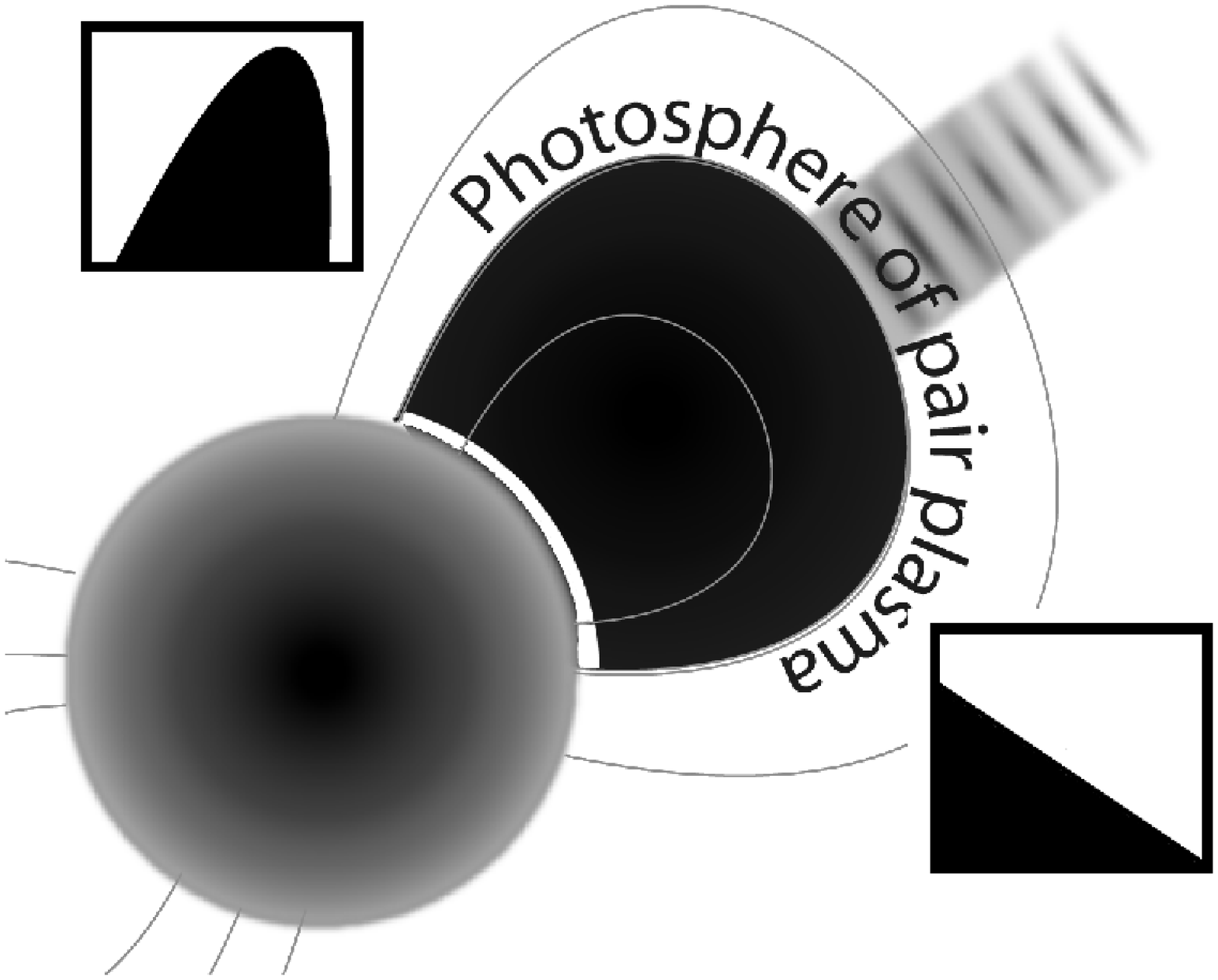}
\caption{A schematic illustration of the formation of the pair fireball
(Not to scale).  The left panel shows the configuration before
sufficient plasma has accumulated to make the vicinity of the neutron
star optically thick.  The right panel depicts the fireball and the
continuing fast-mode breakdown outside the fireball.  The fireball
emits as a blackbody but before the formation of the fireball and
outside of the fireball itself the pairs generate a
distribution of high-energy synchrotron radiation up to $E=2m_e c^2$
(insets).}
\label{fig:schematic}
\end{figure}

As the schematic illustrations in Figure~\ref{fig:schematic}
show, some radiation will escape before the fireball forms and from
outside the fireball.  The fraction of the emission that escapes
before the formation of the fireball is approximately the reciprocal
of the peak optical depth of the fireball, $\sim 10^{-9}$ for the
models considered here.

\begin{figure}
\plottwo{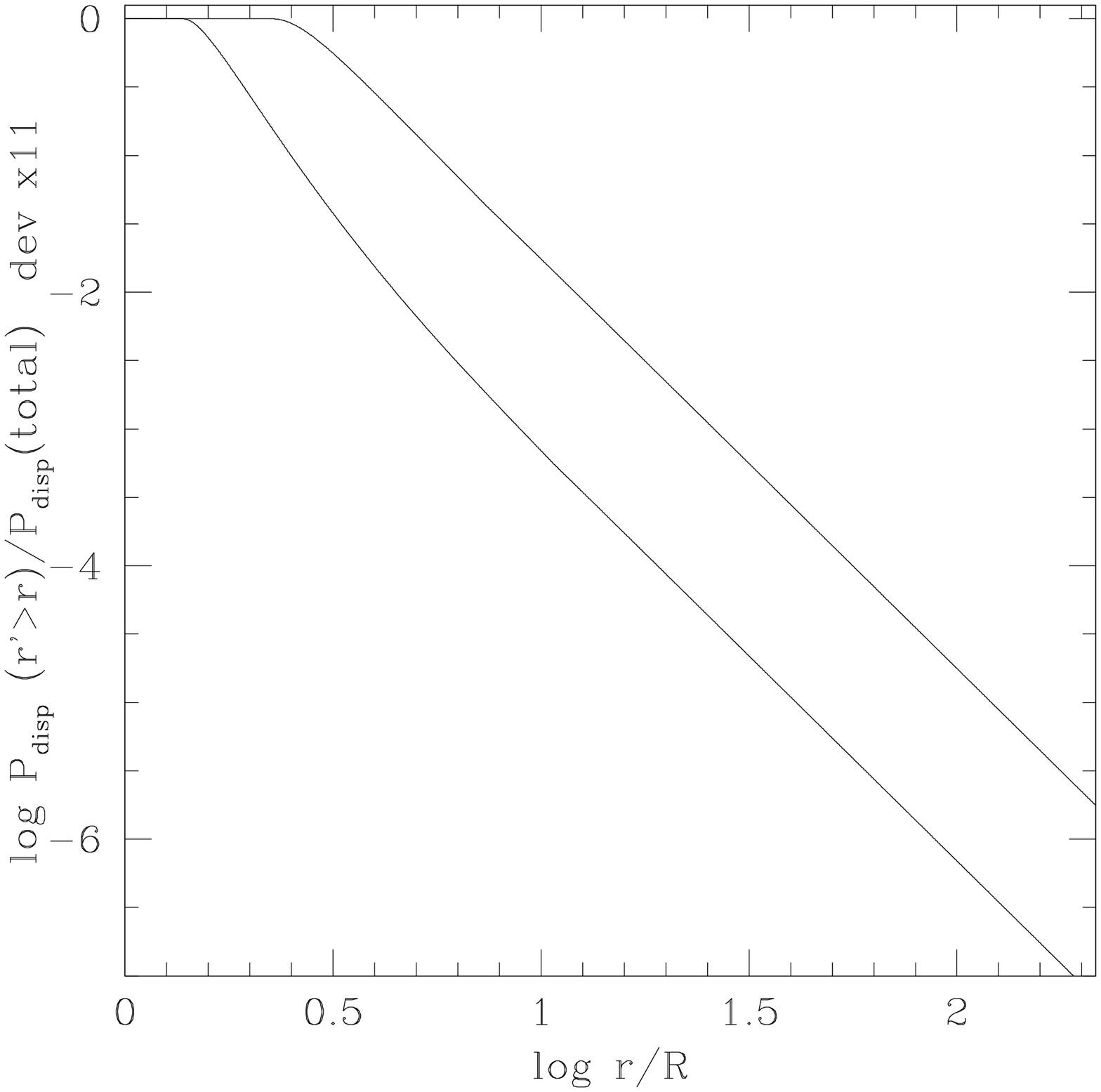}{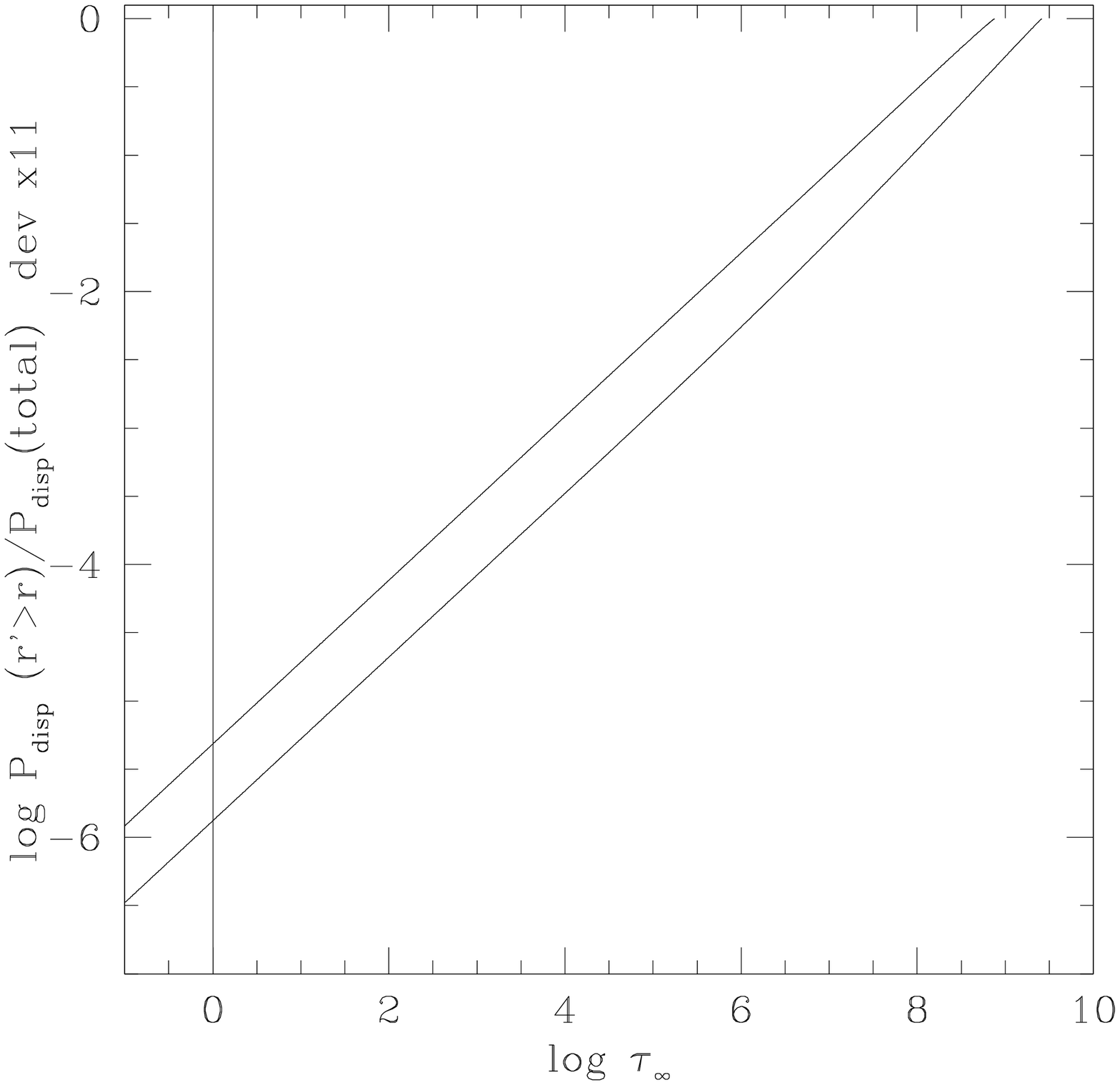}
\caption{The two panels depict the fraction of the pair-production outside
of a given radius or below a particular optical depth. The energy dissipated
below an optical depth of unity will escape to infinity as
synchrotron emission.  The upper curve in both panels is for 
$B_{NS}=10 B_\rmscr{QED}$ and the lower curve is for $B_{NS}=30 B_\rmscr{QED}$.}
\label{fig:outside_emission}
\end{figure}

It turns out that a much larger fraction of the energy is emitted from
outside the fireball.  Figure~\ref{fig:outside_emission} makes this
more quantitative.  In both cases about $10^{-5}$ to $10^{-6}$ (or
$\sim 10^{34}$~erg) of the burst energy emerges as synchrotron emission
extending to $E=2m_e c^2$ (above this energy the photons would produce
pairs in the strong magnetic field of the neutron star).  The schematic
depicts that this emission has a power-law distribution, but this is
merely for convenience.  The calculation of the energy distribution of
this high-energy radiation is beyond the scope of this paper.

The asymptotic scalings show that the emission from outside a
particular radius is proportional to $(k b)^{-2} r^{-3}$.  Combining
this with the result for the location of the photosphere yields that
the fraction of the emission that is produced outside the fireball is
proportional to $(k b)^{-4/5}$.

Our estimates for the structure of the fireball are sensitive to the
global properties of the neutron star magnetic field.  Here, we have
assumed a pure dipole; higher order multipoles would alter some of
our estimates.  For example, if the field included a large quadrupole
contribution, the fireball would be more compact than the estimates
given above.

\section{SGRs and AXPs vs. High-Field Radio Pulsars}

One of the most puzzling aspects of SGRs and AXPs is why they appear
to behave so differently from those radio pulsars with similar timing
characteristics.  Originally, it was believed that SGRs and AXPs might
be distinguished from radio pulsars purely on the basis of the
strengths of their magnetic fields.  However, a number of radio
pulsars are now known that have $B_\rmscr{NS} \gtrsim
B_\rmscr{QED}$ and, in some cases, fields even larger than those of
AXPs.  (Assuming that SGRs and AXPs are isolated objects and are not
accreting.)

Presently, there are five radio pulsars with magnetic fields $B_\rmscr{NS}
\gtrsim B_\rmscr{QED}$ (Camilo et al.  2000; Morris et al. 2002;
McLaughlin et al. 2003a,b); in particular, the magnetic field of
PSR~J1847-0130 is $B_\rmscr{NS}=9.4\times 10^{13}$ G.  It is interesting to
compare PSR~J1847-0130 with 1E2259+586, which is perhaps the most
well-studied AXP.  In many respects, these stars appear to be very
similar.  The period and spin-down rate are 6.71 secs and $1.27\times
10^{-12}$ s s$^{-1}$ for PSR~J1847-0130 (McLaughlin et al. 2003a)
versus 6.98 secs and $4.84\times 10^{-13}$ s s$^{-1}$ for
AXP~1E2259+586 (e.g. Gavriil \& Kaspi 2002).  The implied field
strengths and characteristic ages are $B_\rmscr{NS}=9.4\times 10^{13}$ G and
$\tau_c = 83,000$ years for PSR~J1847-0130 and $B_\rmscr{NS}=
5.9\times 10^{13}$ G
and $\tau_c = 230,000$ years for AXP~1E2259+586.

But, in other ways, these two objects are completely different.
PSR~J1847-0130 is a {\it radio} pulsar, whereas none of the AXPs have
been detected as radio sources.  Bursts of the type analyzed here were
discovered from AXP~1E2259+586 (Kaspi et al. 2003), but {\it no} radio
pulsar has exhibited similar behavior.  These differences cannot be
explained on the basis of magnetic field intensity since, in this
case, the more ``ordinary'' radio pulsar has the {\it larger} inferred
value of $B_\rmscr{NS}$. 
Similarly, the differences cannot be ascribed to the
spin-down rate, because the more ``peculiar'' AXP has the {\it
smaller} measured $\dot{P}$.  Clearly, the true explanation is more
involved.

In what follows, we speculate on reasons that may account for the
contrasting properties of PSR~J1847-0130 and AXP~1E2259+586.
Our scenario is motivated by two other
differences between these objects.  First, AXP~1E2259+586 is
associated with a supernova remnant (CTB 109) with an estimated age of
$\tau_\rmscr{SNR} = 10^4$ years, while PSR~J1847-0130 appears to be
isolated.  Second, the persistent X-ray luminosity of the AXP in the 2
- 10 keV band is much higher: $L_x \sim 10^{34.5} - 10^{35}$ erg/s
compared to an upper limit of $L_x < (2-8) \times 10^{33}$ erg/s for
PSR~J1847-0130 (e.g. McLaughlin et al. 2003a).

One interpretation of these differences is that AXP~1E2259+586 is
considerably younger than PSR~J1847-0130, in spite of its much larger
timing age (230,000 vs. 83,000 years).  This would account at least
partly for its much higher persistent X-ray luminosity (e.g. Heyl \&
Hernquist 1997a,b), as well as its association with a $10^4$ year old
supernova remnant.

In the magnetar model, it is usually stated that the neutron star
is ``magnetically powered,'' in the sense that the magnetic field
represents the dominant available supply of energy.  However, it is
not clear that this is always the case.
The magnetic energy is
\begin{equation}
{\cal M} \, = \, {1\over{8\pi}} \, \int B^2 \, dV \, .
\end{equation}
If the magnetic field throughout the star is uniform, then for
AXP~1E2259, 
\begin{equation}
{\cal M} \, \sim \, 6\times 10^{44} \, R_6^3\, \left ( {B_\rmscr{NS}
\over{5.9\times 10^{13} G}} \right ) ^2 \,\,\,  {\rm ergs},
\end{equation} 
where $R_6$ is the radius of the star in units of $10^6$ cm.  This can
be compared with the available thermal energy of the star, $dU/dT =
C_V$, where $C_V$, the heat capacity (e.g. Shapiro \& Teukolsky 1983),
is proportional to the temperature.  For a uniform density sphere of
non-relativistic neutrons, this can be integrated to give
\begin{equation}
U \, \sim \, 10^{47} \, R_6^2 \, M_{1.4}^{1/3} \,
T_{8.5}^2 \,\,\,\, {\rm ergs},
\end{equation}
where $M_{1.4}$ and $T_{8.5}$ are the mass 
and core temperature of the star in units of $1.4M_\odot$ and
$10^{8.5}$ K, respectively.

Heyl \& Hernquist (1997b) studied the thermal evolution of
ultramagnetized neutron stars and found that if these objects follow
conventional cooling tracks then the core temperature is given by
\begin{equation}
T_9 \, \sim \, t_\rmscr{yr}^{-1/6} \, ,
\end{equation}
where $t_\rmscr{yr}$ is the age in years, and this equation is valid
for $t_\rmscr{yr} \lesssim 10^5$ yrs and $B_\rmscr{NS} \lesssim
10^{15}$ G.  If the true age of AXP~1E2259 is $\sim 10^4$ years, then
$T_{8.5} \approx 0.7$ and from equations (19) and (20), $U \sim 100
{\cal M}$, and so the thermal component represents the largest energy
supply in this star.  This conclusion would not obtain if
unconventional sources of cooling are included.  In this case,
however, the expected thermal emission would not contribute
significantly to the X-ray spectrum of AXP~1E2259 (e.g. Heyl \&
Hernquist 1997a).  Moreover, Heyl \& Kulkarni (1998) have shown that
magnetic field decay for $B \lesssim 10^{15}$ G contributes negligibly
to the reservoir of thermal energy of such a young neutron star.
Consequently, the thermal radiation from AXP 1E2259 is probably not
strongly affected by field decay.

Thus, there is the possibility that at least in the case of AXP~1E2259
the evolution is driven primarily by thermal, rather than magnetic
effects.  In principle, this could distinguish that object from
PSR~J1847-0130 because the radio pulsar may be much older and, hence,
its supply of thermal energy would be depleted relative to the AXP
(Heyl \& Hernquist 1997b).

Blandford, Applegate \& Hernquist (1983) proposed that a vertical heat
flow through the outer layers of a neutron star can amplify a
horizontal seed field by thermoelectric effects.  Whether or not this
mechanism can actually generate the magnetic fields associated with
AXPs is unclear, but the analysis of Blandford et al.  suggests that
the structure of a pre-existing magnetic field can be altered by the
thermal flux.  In particular, the flow of heat may yield a disordered,
evolving pattern of magnetic flux on the surface of the star, whether
or not the crust is seismically active, and regardless of how the
field originated.  The thermal evolution of the star could in
principle lead directly to magnetic activity in the magnetosphere,
producing stresses in the crust and ultimately starquakes. We note
also that the thermoelectric effect studied by Blandford et al. is
most efficient if the outer layers of the star consist of light
elements, as is required for the cooling calculations to match the
observed X-ray luminosity (Heyl \& Hernquist 1997a).

In the examples presented earlier, we assumed a simple dipole geometry
in estimating the opacity for MHD waves to develop shocks.  However,
our model depends not on the dipole component of the field but the
total field near the neutron star, $B_\rmscr{NS}$, which could be
larger and more complicated than a dipole.  A key difference between
AXP~1E2259 and PSR~J1847 is that 1E2259 was identified as an AXP; this
means that its X-ray emission exceeds the spin-down luminosity by many
orders of magnitude.  Regardless of whether this emission is due to
the cooling of the neutron star or magnetic field decay, this heat
emerges through the crust of the neutron star (this does not apply for
accretion models of AXPs), and could drive evolution in the surface
field.

Thus, we propose that the thermal flux in AXP~1E2259 may be
responsible for maintaining magnetic activity in this star, possibly
supplemented by seismic effects, owing to its young age.  This
magnetic activity seeds the mechanism for bursts described in this
paper by generating MHD waves in a continuous manner, some of which
exceed the threshold for shocking owing to their short wavelengths
and/or large amplitudes.  As such a star cools, the surface field
begins to stabilize when the thermal energy is radiated.  The star
becomes more quiescent as the magnetic field becomes mainly dipolar
and the source of MHD waves to trigger bursts is no longer available.
Eventually, it may evolve into an object like PSR~J1847, which
resembles an ordinary radio pulsar, albeit slowly rotating and
unusually strongly magnetized.

In the context of our model, AXP~1E2259 would differ from PSR~J1847
not on the basis of its magnetic field strength or spin-down rate, but
because it is much more magnetically active.  One piece of evidence
that indirectly supports this interpretation is the discrepancy
between the timing age of AXP~1E2259 and the age of its associated
supernova remnant or the cooling age inferred from its X-ray
luminosity.  As we indicated earlier, if the magnetic field structure
of AXP~1E2259 is complex and dynamic, timing estimates for a simple
dipole will not lead to accurate values for the true age of the star.
This is consistent with the notion that the magnetic field of
AXP~1E2259 is disordered, owing to the impact of thermal processes on
the structure of the field.  (This explanation is not unique, however,
as the discrepancy between the timing age and the age of the supernova
remnant can also be accounted for even if the field is dipolar,
provided that the star was born rotating slowly.)

To the extent that AXPs and SGRs are the same type of object, as
suggested by observations, the arguments presented here may be
relevant to this entire subclass of neutron stars.  From 
equations (19) and (20), we can obtain a condition for the
approximate equality of magnetic and thermal energies:
\begin{equation}
B_{15} \, \sim \, {\bar{\rho}}_{15}^{1/6} \, T_{8.5} \, ,
\end{equation}
where ${\bar{\rho}}_{15} = {\bar{\rho}}/10^{15}$ is the mean
density of the star in gm/cm$^3$ and $B_{15} = B_\rmscr{NS}/
10^{15} {\rm G}$.
The largest estimated fields for AXPs and SGRs are $B_\rmscr{NS}
\sim 10^{15}$ G,
and these stars have ages $\sim 10^4$ years, implying core temperatures
$T_{8.5} \sim 0.7$, according to equation (21).  Thus, it is
possible that heat and not magnetism represents the dominant
global energy supply.  Detailed
calculations of the thermal structure
of ultramagnetized neutron
stars (e.g. Heyl \& Hernquist 1998b, 2001), 
combined with modeling of the interplay between magnetic
fields and the heat flux in the outer layers of the stars will
be required to understand the implications of the energetics
more fully.  But, we note that the analysis of Heyl \& Kulkarni
(1998) supports our argument that thermal effects dominate
under the relevant circumstances.

\section{Discussion}

Our theory for the bursts from SGRs and AXPs requires that the
magnetic field in the vicinity of the neutron star exceed the quantum
critical field $B_\rmscr{QED}$.  If the wavelength to amplitude ratio
of fast modes is bounded from below, then there is a magnetic field
below which no bursts can occur.  This cutoff behavior is in contrast
to the model proposed by TD95, in which neutron stars with weaker
fields could, in principle, exhibit weaker bursts.

\subsection{Comparison with TD95}

Many of the details of the TD95 model depend on the strength of the
stellar magnetic field, so processes similar to theirs could produce
bursts from more weakly magnetized neutron stars.  For example, TD95
consider a pair-plasma fireball in a weaker magnetic field in order to
account for Type-II X-ray bursts.  Bursts such as these have not been
detected from any radio pulsar even though the radio pulsars
(including the most strongly magnetized ones) are typically ten to one
hundred times closer to Earth than the SGRs, and instruments such as
HETE-2 routinely detect such events from accreting sources.  If the
Alfv\'enic cascade outlined by TD95 does not operate, the cutoff
behavior of fast-mode breakdown would naturally explain why these
radio pulsars do not burst.

There are other significant differences between the Alfv\'enic cascade
scenario of TD95 and the fast-mode breakdown model presented here.
HH99 argued from the principles of gauge and Lorentz invariance that a
single Alfv\'en wave does not suffer non-linearities in relativistic MHD
(even when coupled to QED); only oppositely directed waves interact.
Consequently, an Alfv\'en wave produced at the surface must propagate
across the magnetosphere of the neutron star and reflect off the
surface before it can interact with itself and instigate a cascade.
Unless the magnetic field is tangled, the wave would have to travel
several stellar radii before the cascade could begin, and furthermore
it would have to reflect efficiently rather than be absorbed.  Because
the cascade requires both waves, the energy of the fireball would be
limited to the energy of the reflected wave.  At each reflection, the
interaction time-scale is limited by the duration of the wave.  If the
stellar magnetic field is sufficiently strong, the fast modes will
break down through the mechanism proposed here within a stellar radius
and dump a large fraction of their energy into pairs, creating a
fireball well before an Alfv\'enic cascade can begin.

Simulating an Alfv\'enic cascade would be amendable to the techniques
used here and in HH98.  The non-linearities in the evolution of
Alfv\'en waves comprise two effects, one magnetohydrodynamic and one
quantum-electrodynamic.  The QED process is essentially the same as
that which operates for the fast modes.  A detailed simulation of an
Alfv\'enic cascade may reveal that it cannot operate within the
confines of a neutron-star magnetosphere unless the QED contribution
is sufficiently strong.  Whether or not such a magnetic cutoff exists
for the Alfv\'en modes is left for future work.

\subsection{Non-thermal emission}

Fast-mode breakdown predicts that high energy emission extending up to
1~MeV should accompany the thermal emission from the pair plasma.
Although the fluence of this high-energy emission is much weaker than
that of the fireball ($\sim 10^{-5}$), its flux may actually be
comparable because it is produced promptly by fast-mode breakdown
itself, rather than by the cooling fireball.  Fast-mode breakdown also
dumps a large number of pairs into the magnetosphere.  If the lifetime
of the pairs is sufficiently long or the smaller-scale and more common
fast modes break down without forming a large fireball (the estimates
of the preceding section assumed that the initial disturbance subtends
a steradian of the stellar surface), fast-mode breakdown can populate
a neutron-star magnetosphere with high-energy electron-positron pairs.
This could fuel a pulsar-like emission process, possibly accounting
for the optical/IR flux from AXPs (Hulleman et al. 2000, 2001; Wang \&
Chakrabarty 2002; Israel et al. 2002, 2003a).  We have not discussed
in detail the properties of the non-thermal component of fast-mode
breakdown emission, but depending on the relative locations of the
optical and infrared emission, one might also explain the relative
variability in the two wave-bands (Kern \& Martin 2002).

\section{Conclusions}

We have presented a model for the origin of bursts from SGRs and AXPs
which we call ``fast-mode breakdown.''  In this theory, certain MHD
waves propagating through the magnetospheres of neutron stars with
sufficiently strong magnetic fields will develop discontinuities in
response to the QED process of vacuum polarization and dissipate as a
pair plasma fireball.  To function, our mechanism requires both a
strong stellar magnetic field, $B_\rmscr{NS} \gtrsim B_\rmscr{QED}
\approx 4.4 \times 10^{13}$~G, as well as an active source of short
wavelength, large amplitude MHD waves.  Owing to the dependence of
fast-mode breakdown on $B_\rmscr{NS}$ and the properties of the wave,
this phenomenon is subject to a cutoff, and will not operate below a
limiting value of $B_\rmscr{NS}$, for a given spectrum of MHD waves.

We have argued that a primary difference between AXPs/SGRs and
high-field radio pulsars may be in their ability to sustain the
production of MHD waves of the appropriate wavelength and amplitude.
On the basis of our model, we propose that only those neutron stars
born with $B_\rmscr{NS} \gtrsim B_\rmscr{QED}$ will exhibit AXP- or
SGR-like behavior when they are young.  If the source of the waves
involves an interplay between thermal and magnetic processes, then
magnetars may evolve into radio pulsars as they age and lose their
thermal energy.  This transition would occur on roughly the cooling
time-scale, which depends on the strength of the magnetic field (Heyl
\& Hernquist 1997b), but is typically $\lesssim 10^5$ years.  This
would explain the absence of old AXPs and SGRs.  Conceivably, this
picture could also account for the lack of AXPs or SGRs with
$B_\rmscr{NS} \gg 10^{15}$ G, if complete magnetic domination
ultimately suppresses the effects that excite the wave activity
required by our model.

Our work suggests interesting directions for future research on
related topics.  As we indicated earlier, the effective Lagrangian we
have employed cannot describe the structure of electromagnetic and MHD
shocks because it assumes that the magnetic field is uniform across an
electron Compton wavelength.  Analyzing this situation will require a
more general approach, such as the proper-time method of Schwinger
(1951).

We have ignored dispersive effects in describing the propagation of
the waves, on the basis of estimates like those given by Adler (1971).
If this approximation is not always appropriate, it raises the
possibility that a competition between wave-steepening from QED and
dispersion from a background plasma could lead to a situation in which
solitary waves (photon torpedoes?) may exist.  The consequences of
such solutions are not currently predicted by our theory.

Here, we have focused on the evolution and break-down of individual
fast-mode waves.  However, as noted by HH98, QED effects will enable
non-linear interactions between Alfv\'en waves in neutron star
magnetospheres that could drive other paths for wave energy to be
dissipated.  The various shocks we have described could also interact
with background plasma, producing accelerated particles with
characteristics similar to ultra high-energy cosmic rays.

Finally, it is of interest to examine the evolution of magnetic fields
in neutron stars, particularly in interaction with the heat flow
through the outer layers of these objects.  Direct simulation of e.g.
thermoelectric coupling between magnetic fields and thermal flux
should be possible, leading to greater understanding of the 
consequences of these processes.

We have suggested how our theory could be tested by, for example,
searching for the high energy tail of emission that should accompany
the mostly thermal spectra of the bursts.  If our model is supported
by such observations, it will clarify the energy source of bursts
from SGRs and AXPs, add to the growing evidence that these objects
have magnetic fields $B_\rmscr{NS} \gtrsim B_\rmscr{QED}$, and
demonstrate the importance of QED effects in astrophysical sources.

\acknowledgments
We thank Roger Blandford and Gary Gibbons for stimulating discussions.
Support for this work was provided by the National Aeronautics and
Space Administration through Chandra Postdoctoral Fellowship Award
Number PF0-10015 issued by the Chandra X-ray Observatory Center, which
is operated by the Smithsonian Astrophysical Observatory for and on
behalf of NASA under contract NAS8-39073, as well as NASA ATP grants
NAG5-12140 and NAG5-13292 and an NSERC Discovery grant.

\bibliographystyle{apj}
\bibliography{mine,physics,ns,gr,sgr}

\end{document}